\documentclass[11pt]{article}

\usepackage{graphicx}

\usepackage{amssymb}
\usepackage{graphicx}

\begin{document}


\title{Polarization kinetics in ferroelectrics \\ with regard
to fluctuations}

\author{J. Kaupu\v{z}s 
\thanks{E--mail: \texttt{kaupuzs@latnet.lv}} \hspace{0ex}, J. Rimshans \\
Institute of Mathematics and Computer Science University of Latvia
\\
29 Rainja Boulevard, LV--1459 Riga, Latvia \\
N.F. Smyth \\
School of Mathematics, University of Edinburgh,\\
King's Buildings, Mayfield Road, Edinburgh, Scotland, U.K., EH9 3JZ.}

\maketitle

\begin{abstract}
Polarization in ferroelectrics, described by the Landau-Ginzburg
Hamiltonian, is considered, based on a multi-dimensional Fokker-Planck
equation.  This formulation describes the time evolution of the probability
distribution function over the polarization field configurations in the
presence of a time dependent external field.  The Fokker-Planck
equation in a Fourier representation is obtained, which can then be solved
numerically for a finite number of modes.  Calculation results
are presented for one and three modes.  These results show the hysteresis
of the mean polarization as well as that of the mean squared gradient of the
polarization.
\end{abstract}

polarization hysteresis.

\vspace{1ex}

\section{Introduction}

The stochastic description of collective phenomena such as phase
transitions is a key theme in solid-state physics.  Problems
of this kind are nontrivial and are usually solved by means of
perturbation theory~\cite{Ma,Justin,K}.  Here the kinetics of
polarization switching in ferroelectrics is studied, taking into
account the spatio-temporal fluctuations of the polarization field
given by the Langevin and multi-dimensional Fokker-Planck
equations~\cite{Parisi,Haken}.  Simplified approaches are well
known~\cite{Liu,Talkner,Wan}, for which the static distribution of the
polarization is found by minimizing the Landau-Ginzburg type free
energy functional.  The dynamics and fluctuations of the polarization
are usually introduced via equations of mean-field type representing
either an approximation or a solution of the appropriate mean-field
model~\cite{Shiino}.  The multi-dimensional Fokker-Planck equation
discussed here allows us to consider unlimitedly many nontrivial degrees
of freedom for the correlated collective fluctuations of the polarization
field, which are lost or roughly treated as uncorrelated in the
mean-field approximation.  This problem has been studied
previously~\cite{K1,K2} by means of the perturbative Feynman diagram
technique.  Such an approach, however, only allows the description of the
ferroelectric response with respect to an infinitely small external
field, rather than polarization reversal (switching) and hysteresis.
The latter requires a non-perturbative treatment~\cite{DM,Klotins,KK} which,
however, is usually done by including some kind of mean-field
approximation. In~\cite{KO1,KO2,KO3} the polarization reversal has been
studied based on the classical nucleation theory. At the initial stage,
the formation of domains of new phase is considered as one--step Markov 
process, in which case the domain size can increase or decrease by one 
unit at a time moment. The obtained in this way distribution over
the sizes of small fluctuating domains (or nuclei) is then used to
describe the later stage of the domain growth and the polarization reversal.
Despite of some simplifications (neglecting nontrivial spatial correlations
as well as complicated  merging and splitting processes) it gives realistic
results~\cite{KO1,KO3}. An explicit consideration of the nucleation
and growth of domains lies also in the basis of more recent 
simulations of ferroelectricity in polycrystals 
and single crystals~\cite{CLKC05,CLKC07}, as well as of combined
ferro-- and piezo--electric effects~\cite{ZB1,ZB2}. These ideas
are applied also to describe experimentally observed features in
specific materials~\cite{IMAMSY,MMSYU}.
In the present work the Fokker-Planck equation in a Fourier
representation, which is suitable for a non-perturbative numerical
approach avoiding approximations of the mean-field type, is derived.  
In principle, it allows to include many nontrivial
degrees of freedom, lost in the known simplified approaches, as regards the 
spatial correlations and fluctuations of the polarization.
In practical calculations of this kind, however, only some fluctuation degrees
of freedom, that is only a few Fourier modes, can be implemented due to
computation time and memory constraints.  Nevertheless, such limited
calculations can reproduce some important qualitative features and provide
test examples to verify possible approximations which would allow the
treatment of realistic numbers of modes.

\section{Basic equations}

We consider a ferroelectric with the Landau-Ginzburg Hamiltonian
\begin{equation}
H = \int \left( \frac{\alpha}{2} P^2({\bf x}) + \frac{\beta}{4} P^4({\bf x})
+ \frac{c}{2} ( \nabla P({\bf x}))^2 - \lambda({\bf x},t) P({\bf x},t)
\right) d {\bf x} \;,
\label{eq:H}
\end{equation}
where $P({\bf x},t)$ is the local polarization and $\lambda({\bf x},t)$
is the time-dependent external field.  The only configurations of the
polarization which are allowed are those corresponding to the cut-off
$k<\Lambda$ in Fourier space with $\Lambda = \pi/a$, where $a$ is the
lattice constant.  The Hamiltonian~(\ref{eq:H}) can be approximated by a sum
over discrete cells, where the size of one cell can be larger than
the lattice constant.  Such cells are small domains with almost constant
polarization.  Thus the Hamiltonian of a system with volume $V$ is
\begin{equation}
H = \Delta V \sum\limits_{\bf x}
\left( \frac{\alpha}{2} P^2({\bf x}) + \frac{\beta}{4} P^4({\bf x})
+ \frac{c}{2} ( \nabla P({\bf x}))^2 - \lambda({\bf x},t) P({\bf x},t)
\right)  \;,
\label{eq:Hd}
\end{equation}
where $\Delta V=V/N$ is the volume of one cell and the coordinates of the
centers of the cells are given by the set of discrete $d$-dimensional vectors
$\bf x \in \mbox{R}^d$.  The stochastic dynamics of the system is described
by the Langevin equation
\begin{equation}
\dot{P}({\bf x},t) = -\gamma \frac{\partial H}{\partial P({\bf x},t)}
+ \xi({\bf x},t) \;,
\end{equation}
where $\xi({\bf x},t)$ is white noise, i.e.
\begin{equation}
\langle \xi({\bf x},t) \xi({\bf x}',t') \rangle
= 2 \gamma \theta \, \delta_{{\bf x},{\bf x}'} \delta(t-t') \;.
\end{equation}
In the case of Gaussian white noise, the probability distribution
function
$$
f \left( P \left({\bf x}_1 \right), P \left({\bf x}_2 \right), \cdots,
P \left({\bf x}_N \right), t \right)
$$
is given by the Fokker-Planck equation~\cite{Haken}
\begin{equation}
\frac{1}{\gamma} \frac{\partial f}{\partial t} =
\sum\limits_{\bf x} \frac{\partial}{\partial P({\bf x})}
\left( \frac{\partial H}{\partial P({\bf x})} \, f +
\theta \frac{\partial f}{\partial P({\bf x})} \right) \;.
\end{equation}
At equilibrium the flux vanishes, which corresponds to the
Boltzmann distribution $f \propto \exp(-H/\theta)$ with $\theta=k_BT$.

Assuming periodic boundary conditions, we consider the discrete Fourier
transform
\begin{eqnarray}
P({\bf x}) &=& N^{-1/2} \sum\limits_{\bf k} P_{\bf k} e^{i{\bf kx}}
\label{eq:Ftr} \\
P_{\bf k} &=& N^{-1/2} \sum\limits_{\bf x} P({\bf x}) e^{-i{\bf kx}} \;.
\end{eqnarray}
The Fourier amplitudes are the complex numbers $P_{\bf k} = P'_{\bf k}
+i P''_{\bf k}$.  Since $P({\bf x})$ is real, $P'_{-\bf k}=P'_{\bf k}$
and  $P''_{-\bf k}=-P''_{\bf k}$ hold.  It is supposed that the total number
of modes $N$ is an odd number.  This means that there is a mode with
${\bf k}={\bf 0}$ and modes with $\pm {\bf k}_1, \pm {\bf k}_2,
\ldots, \pm {\bf k}_m$, where $m=(N-1)/2$ is the number of independent
non-zero modes.

The Fokker-Planck equation for the probability distribution function
\begin{equation}
f = f \left( P_{\bf 0}, P'_{{\bf k}_1},P'_{{\bf k}_2}, \ldots,
P'_{{\bf k}_m}, P''_{{\bf k}_1},P''_{{\bf k}_2}, \ldots, P''_{{\bf k}_m},
t \right)
\end{equation}
is then
\begin{eqnarray}
\frac{1}{\gamma} \frac{\partial f}{\partial t} &=&
\sum\limits_{{\bf k} \in \Omega} \frac{\partial}{\partial P'_{\bf k}}
\left\{ \frac{1}{2} \left(1+\delta_{{\bf k},{\bf 0}} \right)
\left[ \frac{\partial H}{\partial P'_{\bf k}} f + \theta
\frac{\partial f}{\partial P'_{\bf k}} \right] \right\}
\nonumber \\
&+& \sum\limits_{{\bf k} \in \bar\Omega} \frac{\partial}{\partial P''_{\bf k}}
\left\{ \frac{1}{2}
\left[ \frac{\partial H}{\partial P''_{\bf k}} f + \theta
\frac{\partial f}{\partial P''_{\bf k}} \right] \right\} \;,
\label{eq:FPF}
\end{eqnarray}
where $P'_{\bf 0} \equiv P_{\bf 0}$, $\bar\Omega$ is the set of $m$
independent non-zero wave vectors and $\Omega$ includes ${\bf k}={\bf 0}$.
Here the Fourier transformed Hamiltonian is given by
\begin{eqnarray}
H &=& \Delta V \left( \frac{1}{2} \sum\limits_{\bf k}
\left( \alpha +c {\bf k}^2 \right)
\mid P_{\bf k} \mid^2 + \frac{\beta}{4} N^{-1}
\sum\limits_{{\bf k}_1+{\bf k}_2+{\bf k}_3+{\bf k}_4={\bf 0}}
P_{{\bf k}_1} P_{{\bf k}_2} P_{{\bf k}_3} P_{{\bf k}_4} \right. \nonumber \\
&&\hspace{8ex} - \left. \sum\limits_{\bf k} \lambda_{-\bf k}(t) P_{\bf k} \right) \;.
\label{eq:HF}
\end{eqnarray}
Some of the variables in Eq.~(\ref{eq:HF}) are dependent according to
$P'_{-\bf k} \equiv P'_{\bf k}$ and $P''_{-\bf k} \equiv -P''_{\bf k}$.
Here $\lambda_{\bf k}(t) = \lambda'_{\bf k}(t) + i \lambda''_{\bf k}(t)$
is the Fourier transform of $\lambda({\bf x},t)$.  The Fokker-Planck
equation~(\ref{eq:FPF}) can be written explicitly as
\begin{eqnarray}
\frac{1}{\gamma} \frac{\partial f}{\partial t} &=&
\sum\limits_{{\bf k} \in \Omega} \frac{\partial}{\partial P'_{\bf k}}
\left\{ \Delta V \, f \left[ \left( \alpha+c {\bf k}^2 \right) P'_{\bf k}
+ \beta S'_{\bf k} -\lambda'_{\bf k}(t) \right]
+ \frac{\theta}{2} \left(1+ \delta_{{\bf k},{\bf 0}} \right)
\frac{\partial f}{\partial P'_{\bf k}} \right\}
\nonumber \\
&+& \sum\limits_{{\bf k} \in \bar\Omega} \frac{\partial}{\partial P''_{\bf k}}
\left\{ \Delta V \, f \left[ \left( \alpha+c {\bf k}^2 \right) P''_{\bf k}
+ \beta S''_{\bf k} -\lambda''_{\bf k}(t) \right]
+\frac{\theta}{2} \frac{\partial f}{\partial P''_{\bf k}} \right\} \;,
\label{eq:FPFex}
\end{eqnarray}
where
\begin{eqnarray}
S'_{\bf k} &=& N^{-1} \sum\limits_{{\bf k}_1+{\bf k}_2+{\bf k}_3={\bf k}}
\left\{ P'_{{\bf k}_1} P'_{{\bf k}_2} P'_{{\bf k}_3}
-3 P'_{{\bf k}_1} P''_{{\bf k}_2} P''_{{\bf k}_3} \right\} \;, \\
S''_{\bf k} &=& N^{-1} \sum\limits_{{\bf k}_1+{\bf k}_2+{\bf k}_3={\bf k}}
\left\{ -P''_{{\bf k}_1} P''_{{\bf k}_2} P''_{{\bf k}_3}
+3 P''_{{\bf k}_1} P'_{{\bf k}_2} P'_{{\bf k}_3} \right\} \;.
\end{eqnarray}

\section{Spatially homogeneous case}

The simplest case is a spatially homogeneous polarization for which only the
${\bf k}={\bf 0}$ mode is retained in~(\ref{eq:FPFex}) with a spatially
homogeneous external field $\lambda({\bf x},t)=\lambda_{\bf 0}(t)
=A \sin (\omega t)$.  In this case we have
\begin{equation}
\frac{1}{\gamma} \frac{\partial f}{\partial t} =
\frac{\partial}{\partial P_{\bf 0}} \left\{ V \, f \left[
\alpha P_{\bf 0} + \beta P_{\bf 0}^3 - A \sin (\omega t) \right]
+ \theta \frac{\partial f}{\partial P_{\bf 0}} \right\} \;.
\label{eq:FPh}
\end{equation}
The mean value of the zero-th Fourier amplitude
\begin{equation}
\bar P_{\bf 0} = \int\limits_{-\infty}^{\infty} P_{\bf 0} f
\left( P_{\bf 0},t \right) d P_{\bf 0}
\end{equation}
is the mean polarization $\bar P$ in this case, as follows from
(\ref{eq:Ftr}).

\section{Quasi one-dimensional case with homogeneous \\ external field}
\label{sec:quasione}

We shall further consider a quasi one-dimensional case, for which
a three-dimen\-sional ferroelectric sample is stretched out in the
$x$-direction, i.e. $L_x \gg L_y$ and $L_x \gg L_z$ hold for the
linear sizes.  In this case we assume that the polarization, as
well as the external field, depend only on the $x$ coordinate.
This means that the wave vectors also have only one non-vanishing
component, which is a scalar quantity $k_n = \left( 2 \pi/L_x
\right) \cdot n$, where $n= 0,\pm 1,\pm 2,\ldots, \pm m$.  From
now on, we shall omit the vector notation in this quasi
one-dimensional case.

As a first step, we shall include only one ($m=1$) independent
non-zero wave vector $k_L=2\pi/L_x$ (in total $N=3$ wave vectors
$k= -k_L,0,k_L$) and a homogeneous external field
\begin{equation}
\lambda(x,t)= \frac{1}{\sqrt{3}} \lambda_0(t)=
\frac{1}{\sqrt{3}} A \sin(\omega t).
\end{equation}
Furthermore, we shall assume that the probability distribution
function in real space is translation invariant at the initial
time.  Due to the translational symmetry of the model, translation
invariance then also holds at all later times.  In the Fourier
representation, this means that the probability distribution
function depends on the modulus of $P_{{k_1}}$, but not on its
phase.  Thus we have
\begin{equation}
f = f \left( P_0, P'_{{k_1}}, P''_{{k_1}},t \right) = \frac{\hat f
\left( P_0, \mid P_{{k_1}} \mid, t \right)} { 2 \pi \mid P_{{k_1}}
\mid} \;,
\end{equation}
where $\hat f \left( P_0, \mid P_{{k_1}} \mid, t \right)$ is the
probability density in the $\left( P_0, \mid P_{{k_1}} \mid
\right)$ space and which obeys the Fokker-Planck equation
\begin{eqnarray}
\frac{1}{\gamma} \frac{\partial \hat f}{\partial t} &=&
\frac{\partial}{\partial P_0} \left\{ \Delta V \, \hat f \left[
\alpha P_0 + \beta \left( \frac{1}{3} P_0^3 + 2 P_0 \mid P_{{k_1}}
\mid^2 \right) - A \sin(\omega t) \right] + \theta \frac{\partial
\hat f}
{\partial P_0} \right\} \nonumber \\
&+& \frac{\partial}{\partial \mid P_{{k_1}} \mid} \left\{ \Delta V
\, \hat f \left[ \left( \alpha+ck_1^2 \right) \mid P_{{k_1}} \mid
+\beta \left( \mid P_{{k_1}} \mid^3 + P_0^2 \mid P_{{k_1}} \mid
\right)
\right] \right. \nonumber \\
&+& \left. \frac{\theta}{2} \left[ \frac{\partial \hat f}
{\partial \mid P_{{k_1}} \mid} - \frac{ \hat f}{\mid P_{{k_1}}
\mid} \right] \right\} \;.
\end{eqnarray}
Since $f$ is finite, $\hat f$ vanishes at $\mid P_{{k_1}} \mid =
0$. The physical boundary conditions correspond to zero flux at
the boundaries $P_0= \pm \infty$, $\mid P_{{k_1}} \mid=0$ and
$\mid P_{{k_1}} \mid=\infty$. An appropriate initial condition has
to be chosen which fulfils these relations, e.g. $\hat f \left(
P_0, \mid P_{{k_1}} \mid, 0 \right) \propto \mid P_{{k_1}} \mid
\exp \left( -a_0 P_0^2 -a_1 \mid P_{{k_1}} \mid^2 \right)$.

\section{Numerical results}

The Fokker-Planck Eq.~(\ref{eq:FPFex}) has been solved numerically
in the quasi-one dimensional case for which the wave vector has
only one component $k \equiv k_x$, in the spatially homogeneous
approximation with only the $n=0$ $\left(k_x=0\right)$ mode (in
total $N=1$ modes), as well as in the next higher order
approximation for which the spatial distribution of the
polarization field is taken into account by including $n=0,\mp 1$
(in total $N=3$ modes) with $k_n = k_L \, n$.

At first, rewriting the Eq.~(\ref{eq:FPFex}) in the following
form:
\begin{eqnarray}
&& \frac{1}{\gamma} \frac{\partial f}{\partial t}=
\frac{\partial}{\partial P'_{ k_0}} J_0^{'} +
\frac{\partial}{\partial P'_{ k_1}} J_1^{'} +
\frac{\partial}{\partial P''_{ k_1}} J_{1}^{''}, \label{eq:FPN1}
\end{eqnarray}
where (for $\alpha<0$):
\begin{eqnarray}
J_n^{\diamond} =  \Delta V \,  \left[ \left( |\alpha|-c { k_n}^2
\right) P^{\diamond}_{ k_n} - \beta S^{\diamond}_{ k_n}
+\lambda^{\diamond}_{k_n}(t)\right]f - r_n^{\diamond}
\frac{\theta}{2} \frac{\partial f}{\partial P^{\diamond}_{ k_n}},
\label{eq:FPN2}
\end{eqnarray}
\begin{equation}
r_n^{\diamond}=\left\{
\begin{array}{r}
1+ \delta_{k_n,0} ; \diamond =' \hspace{0.5ex} \\
 1 ; \diamond = ''
\end{array}
\right., \quad n=0,1.
\end{equation}

Then, by using a special exponential-type substitution for the
distribution function:
\begin{eqnarray}
f = W_{k_n}^{\diamond} \exp \left\{ \int_{p_{0}}^{P^{\diamond}_{
k_n}} \frac{ 2 \Delta V  }{r_n^{\diamond}\, \theta }\left[ \left(
|\alpha|-c { k_n}^2 \right) p - \beta S^{\diamond}_{ k_n}
+\lambda^{\diamond}_{k_n}(t)\right] dp \right\},
\end{eqnarray}
where $W_{k_n}^{\diamond}$ is a normalization function and  $p_0$
is a real number (similarly as in \cite{PolskyRimshans1986}, it
can be shown that  $W_{k_n}^{\diamond}$ and $p_0$ should not
affect final coefficients of a difference scheme), a monotone,
exponential difference scheme has been developed:
\begin{eqnarray}
&&\left(  \Lambda \left( \eta \right) f^{l+1} \right)_{\bf i} =
\sum_{n=0}^{1} \frac{1}{h_{i_n^{'}}^*}\left(A_{\bf i}^n\right)^{'}
f^{l+1}_{{\bf i}-{\bf e}_n^{'}} +
\frac{1}{h_{i_n^{'}}^*}\left(B_{\bf i}^n\right)^{'} f^{l+1}_{{\bf
i}+{\bf e}_n^{'}} - \left(Q_{{\bf i}}^n\right)^{'}
f_{{\bf i}}^{l+1} + \nonumber \\
&&\sum_{n=-1}^{-1} \frac{1}{h_{i_n^{''}}^*}\left(A_{{\bf
i}}^n\right)^{''} f^{l+1}_{{\bf i}-{\bf e}_n^{''}} +
\frac{1}{h_{i_n^{''}}^*}\left(B_{{\bf i}}^n\right)^{''}
f^{l+1}_{{\bf i}+{\bf e}_n^{''}} - \left(Q_{{\bf i}}^n\right)^{''}
f_{{\bf i}}^{l+1}=\frac{1}{\gamma} \frac{ f_{{\bf
i}}^{l+1}-f_{{\bf i}}^{l}
}{\tau_{l+1}}, \label{eq:FPN3}\\
&& \quad \quad \quad \quad \quad \quad \quad \quad \quad \quad
\quad  0<i_n^{\diamond}<I_n^{\diamond}, \nonumber
\end{eqnarray}
with coefficients:
\begin{eqnarray}
&&\left(Q_{{\bf
i}}^n\right)^{\diamond}=\frac{1}{h^{*}_{i_n^{\diamond}}}
\left(A_{{\bf i}+{\bf e}_n^{\diamond}}^n\right)^{\diamond} +
\frac{1}{h^{*}_{i_n^{\diamond}}} \left(B_{{\bf i}-{\bf
e_n^{\diamond}}}^n\right)^{\diamond},
\label{eq:Q}\\
&&\left(A_{{\bf i}}^n\right)^{\diamond}  = \frac{\theta \,
r_n^\diamond}{2} \left(\eta\right)_{{\bf i}-1/2 {\bf
e_n^{\diamond}}} \frac { 1 }{ h_{i_{n}^{\diamond}} \left( \exp
\left(\left(\eta\right)_
{{\bf i}-1/2 {\bf e_n^{\diamond}}} \right)  -1 \right)   } ,  \label{eq:A}\\
&&\left(B_{{\bf i}}^n\right)^{\diamond}  =  \frac{\theta \,
r_n^\diamond}{2} \left(\eta\right)_{{\bf i}+1/2 {\bf
e_n^{\diamond}}} \frac { \exp \left(\left(\eta\right)_ {{\bf
i}+1/2 {\bf e_n^{\diamond}}} \right) }{ h_{i_{n}^{\diamond}+1}
\left( \exp \left(\left(\eta\right)_
{{\bf i}+1/2 {\bf e_n^{\diamond}}} \right)  -1 \right)   } ,  \label{eq:B}\\
&&\left(\eta\right)_ {{\bf i}-1/2 {\bf e_n^{\diamond}}}= \frac{2
\Delta V }{ \theta \, r_n^{\diamond}}\left( \left( - | \alpha |+c
{ k_n}^2 \right) P_{ k_n}^{\diamond} + \beta S_{ k_n}^{\diamond}-
\lambda_{k_n}(t_{l+1}) \right)_ {{\bf i}-1/2 {\bf
e_n^{\diamond}}}h_{i_n^{\diamond}}, \label{eq:Etha}
\end{eqnarray}
where
${\bf i}=(i_0,i_1^{'},i_1^{''})$ and  ${\bf e}_n^{\diamond}$ is
unit vector in the $i_n^{\diamond}$ direction, $l$ is the time
index, $h$ is the polarization space step and $\tau$ is the time
step.

Elaborated difference scheme (\ref{eq:FPN3}) - (\ref{eq:Etha}) has
a first order truncation error in time and a second order
truncation error in space, when $| \eta | \rightarrow 0$. In the
limit $| \eta | \rightarrow \infty$, so that transport is
advection dominated, when $\theta \rightarrow 0 $, the difference
scheme (\ref{eq:FPN3}) - (\ref{eq:Etha}) has first order
truncation errors in time and space.

For one dimensional case, when $N=1$, difference scheme
(\ref{eq:FPN3}) - (\ref{eq:Etha}) coincides with elaborated in
\cite{KaupRim2005}. As it was shown \cite{KaupRim2005}, for $|
\eta | \rightarrow 0$ and $| \eta | \rightarrow \infty$ one
dimensional exponential scheme becomes absolutely stable. Here in
order to control the precision (stability) of the calculations we
adopted a linear prognosis criterion. Accordingly, we required
that the linear prognosis at each time step should not deviate
from the numerical solution to a given precision, which varied
from $\varepsilon_l =2 \cdot 10^{-4}$ to $\varepsilon_f =5 \cdot
10^{-4}$.  If this precision criterion was met, then the time step
was kept constant. The time step was increased if the linear
prognosis deviation was less than $\varepsilon_l$, and it was
decreased if the linear prognosis deviation was greater than
$\varepsilon_f$.

Elaborated absolute monotone scheme (\ref{eq:FPN3}) -
(\ref{eq:Etha}) for the Fokker--Planck Eq.~(\ref{eq:FPFex}) has
been implemented on parallel processing for one and three
polarization space dimensions, when $N=1$ and $N=3$. A parallel
code version of program is written by implementing MPI programming
technology in FORTRAN and using ScaLAPACK (SUN S3L) linear algebra
solver.

In contrast to the quasi-one dimensional case discussed in
Sec.~\ref{sec:quasione}, here we consider homogeneous as well as
non-homogeneous external fields based on (\ref{eq:FPFex}) as a
generic equation.  This equation reduces to (\ref{eq:FPh}) for
$N=1$.  In the case for which the field is spatially homogeneous
it is given by $\lambda_0(t) = A \sin( \omega t) = A' \sqrt{N}$
$\sin( \omega t)$, the other Fourier components being zero,
whereas the non-homogeneous field is chosen such that
$\lambda_0(t) = \lambda'_{k_1}(t) = \lambda''_{k_1}(t) = A'
\sqrt{N} \sin( \omega t)$.  In the coordinate representation, this
corresponds to $\lambda(x,t) = A' \sin( \omega t)$ in the
homogeneous case and to 
$$\lambda(x,t) = A' \sin( \omega t) \\
\left[ 1 + 2 \sqrt{2}\cos(k_1 x+ \pi/4) \right]$$
in the non-homogeneous case.  Note that, according to (\ref{eq:Ftr}), the
mean polarization is given by $\bar P = \bar P_0/\sqrt{N}$.  We
shall also compute the mean-squared gradient of the polarization
$\overline{(\nabla P)^2}$, which is given by
\begin{equation}
\overline{{(\nabla P)}^2} = N^{-1} \sum\limits_{\bf k} {\bf k}^2
\overline{{\left| P_{\bf k} \right|}^2} = N^{-1} \sum\limits_{\bf k} k^2
\left( \overline{{P'_{\bf k}}^2} + \overline{{P''_{\bf k}}^2} \right)
\label{eq:grad}
\end{equation}
in general.  In our quasi one-dimensional case the sum over ${\bf
k}$ reduces to the sum over $k= -k_L,0,k_L$.

In the Eq. (\ref{eq:FPFex}) expressions for $S$ were given by
$S_0=P_0^3$ for $N=1$, and for $N=3$ these are:
\begin{eqnarray}
S_0&=&\frac{1}{3}\left(P_0^{'}\right)^3+2
P_0^{'}\left(\left(P_{k_1}^{'}\right)^2+
\left(P_{k_1}^{''}\right)^2\right), \\
S_{k_1}^{'}&=&\left(P_{k_1}^{'}\right)^3+
P_{k_1}^{'}\left(\left(P_{0}^{'}\right)^2+
\left(P_{k_1}^{''}\right)^2\right), \\
S_{k_1}^{''}&=&\left(P_{k_1}^{''}\right)^3+
P_{k_1}^{''}\left(\left(P_{0}^{'}\right)^2+
\left(P_{k_1}^{'}\right)^2\right).
\end{eqnarray}

The numerical calculations have been performed within the interval
$[-3.5,3.5]$ for each Fourier amplitude ($P_0$ for $N=1$ and
$P_0$, $P'_{k_1}$, $P''_{k_1}$ for $N=3$) with the boundary
condition $f=0$ at the borders of this domain.  The parameter
values are $\gamma=V=\beta=1$, $L_x = 7$, $\alpha=-1$,
$\theta=0.05$, $A'=0.5$ (i.e. $A= 0.5 \sqrt{N}$) and
$\omega=10^{-3}$.  The initial condition
\begin{equation}
f \left( P_0,0 \right) = \frac{1}{\sqrt{2 \pi} \sigma}
\exp \left( - \frac{ \left( P_0 - \tilde P \right)^2}{2 \sigma^2} \right)
\end{equation}
has been used in the spatially homogeneous ($N=1$) case and has been
modified to
\begin{equation}
f \left( P_0,P'_{k_1},P''_{k_1},0 \right) = \frac{1}{(2 \pi)^{3/2}
\sigma^3} \exp \left( - \frac{ \left( P_0 - \tilde P \right)^2 +
{P'_{k_1}}^2 + {P''_{k_1}}^2}{2 \sigma^2} \right)
\end{equation}
in the other case of $N=3$ modes, with $\sigma=0.3$, $\tilde P = -1$
for $P_0<0$ and $\tilde P = 1$ for $P_0>0$.

\begin{figure}
\begin{center}
\includegraphics[scale=0.4]{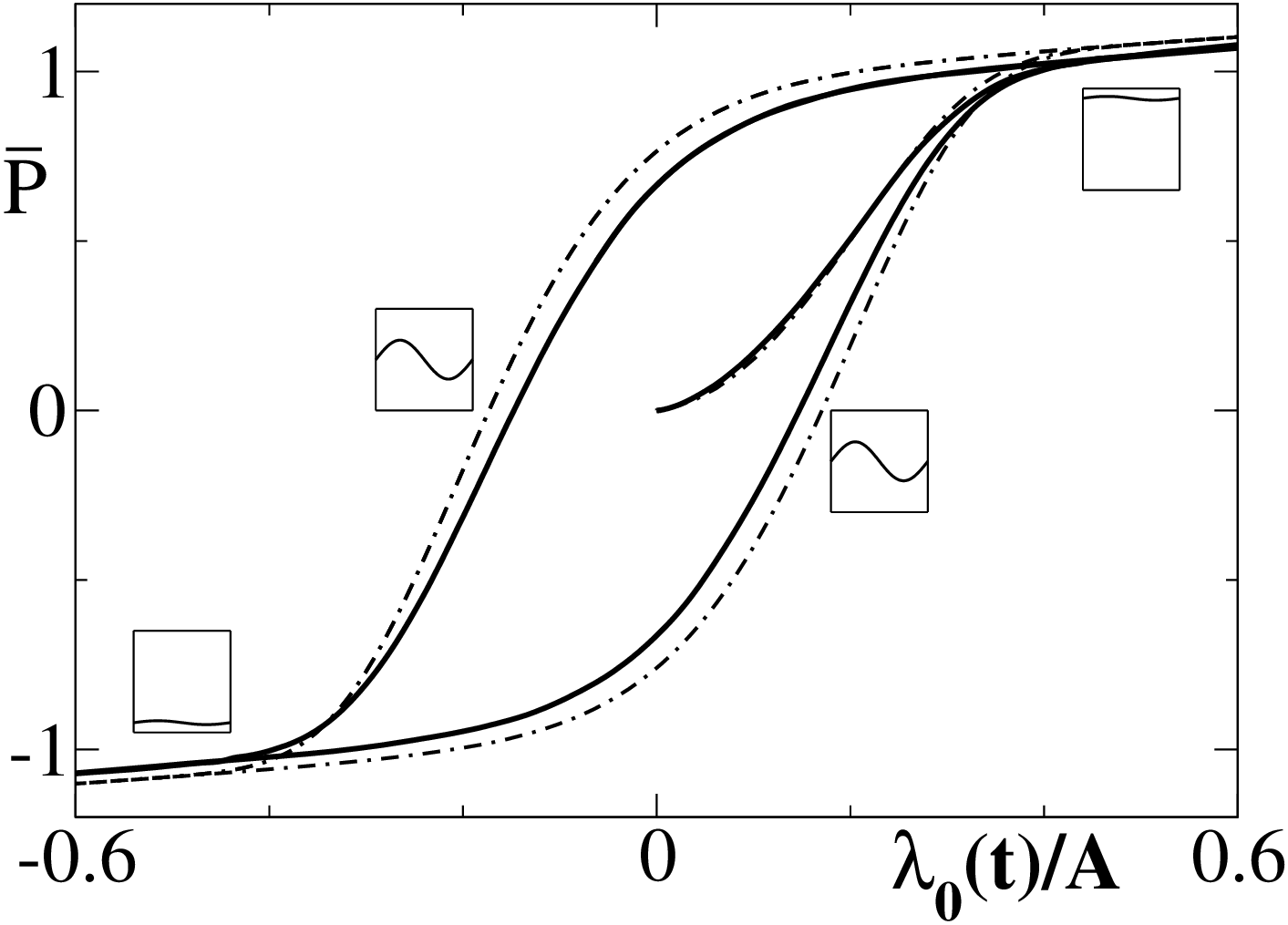}
\end{center}
\caption{The polarization hysteresis: the mean polarization
$\bar P$ vs.\ normalized external field $\lambda_0(t)/A$
calculated numerically for the parameter values $\gamma=V=\beta=1$,
$\alpha=-1$, $\theta=0.05$, $A=0.5 \sqrt{N}$ and $\omega=10^{-3}$.
The solid and dot-dashed lines represent the numerical results
for $N=3$ and $N=1$ Fourier modes, respectively.
The insertions illustrate possible spatial distributions of the 
polarization at different locations on the hysteresis loop.
}
\label{hist}
\end{figure}

The calculated mean polarization in the coordinate space $\bar P =
\bar P_0/\sqrt{N}$, which depends on the homogeneous external
field represented by $\lambda_0(t)$, forms a hysteresis loop, as
shown in Fig.~\ref{hist}, where the results for $N=1$ (dot-dashed
curve) and $N=3$ (solid curve) are compared. In the second case
the spatial distribution of the polarization, generally, is 
nonhomogeneous. The insertions in Fig.~\ref{hist} 
illustrate qualitatively some of possible or typical polarization distributions
at different locations on the hysteresis loop. 
As can be seen, the inclusion of $N=3$ modes only slightly changes the
behaviour of the mean polarization as compared with the homogeneous
approximation with only the zero mode $k=0$ included ($N=1$ case).  The
hysteresis loop becomes narrower for $N=3$.  This is clear physically,
since $N=3$ additional degrees of freedom for the fluctuations in the
polarization make reversal of the polarization easier.
This phenomenon is related to the known Landauer paradox. Namely, the
observed coercive field in real ferroelectrics is much smaller than
that calculated assuming a spatially homogeneous switching of the
polarization. It is well known that switching in ferroelectrics
is a process driven by the nucleation and growth of domains
of different polarization~\cite{KO1,KO3,CLKC07,IMAMSY}. Hence, the above mentioned effect of 
narrowing the hysteresis loop should be even much stronger for
appropriate system parameters and larger number of Fourier modes $N$,
allowing to describe the domain structure realistically.

\begin{figure}
\begin{center}
\includegraphics[scale=0.4]{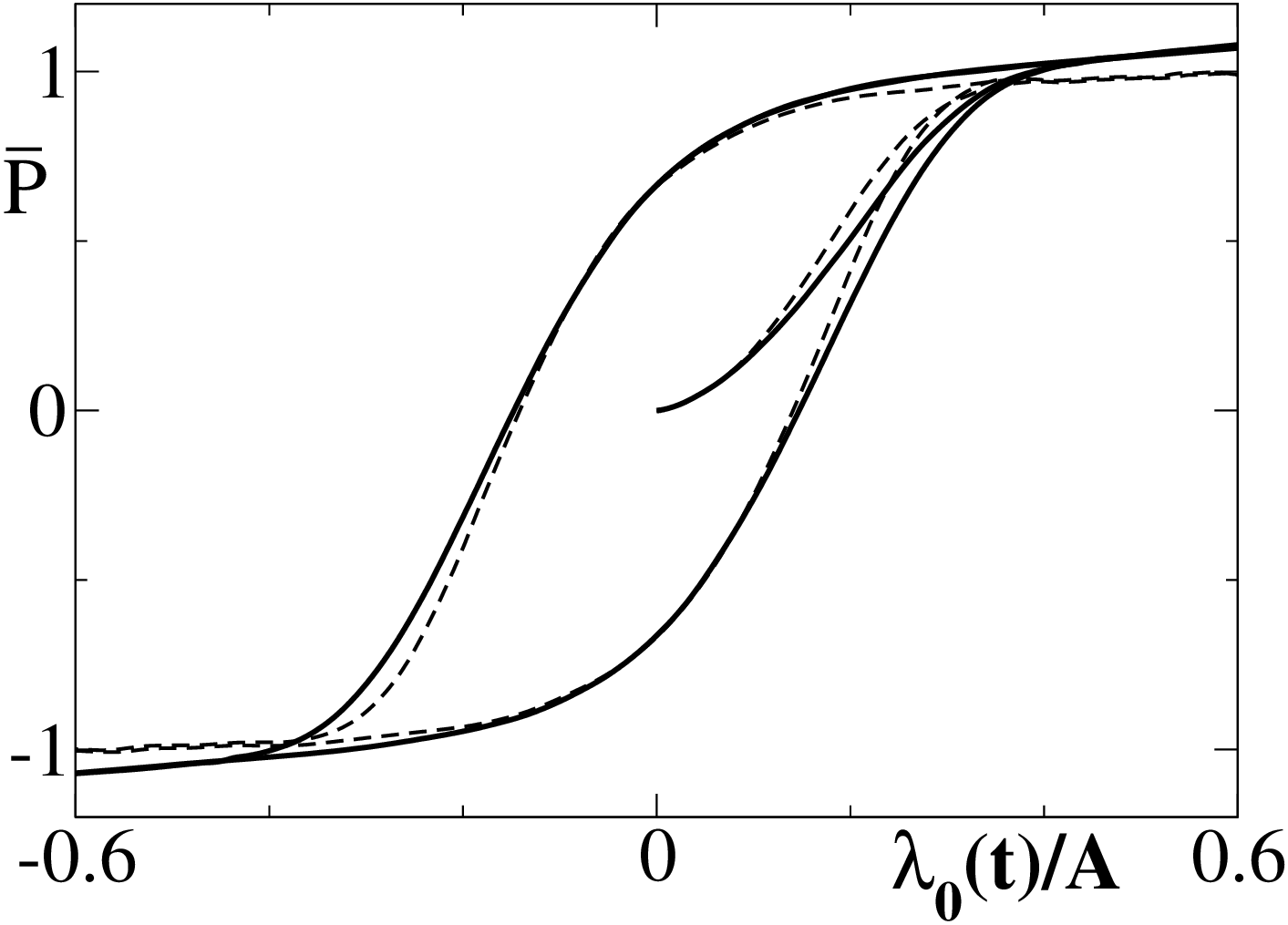}
\end{center}
\caption{The polarization hysteresis: the mean polarization
$\bar P$ vs.\ $\lambda_0(t)/A$ calculated numerically for $N=3$ Fourier
modes.  The solid and the dashed lines represent the results
for a homogeneous and non-homogeneous external field, respectively.
}
\label{hist_h_nh}
\end{figure}

The effect of non-homogeneity of the external field on the polarization
hysteresis can be seen in Fig.~\ref{hist_h_nh}, where numerical
results for the homogeneous (solid curve) and the above discussed
non-homogeneous (dashed curve) fields are compared in the three mode
($N=3$) approximation.
The maximum values of $\mid \bar P \mid$ are somewhat smaller for the
non-homogeneous field.  This is related to the fact that the non-homogeneous
field
\begin{equation}
\lambda(x,t) = A' \sin( \omega t) \left[ 1 + 2 \sqrt{2}
\cos(k_1 x+ \pi/4) \right]
\end{equation}
changes sign depending on the coordinate $x$ in such a way that it
is positive in $61.5\%$ of the volume and negative in the rest, or
vice versa.  This means that the polarization, with a large
probability, has an opposite sign in the smaller part of the volume, which
reduces the modulus of the mean polarization.

\begin{figure}
\vspace{5mm}
\begin{center}
\includegraphics[scale=0.4]{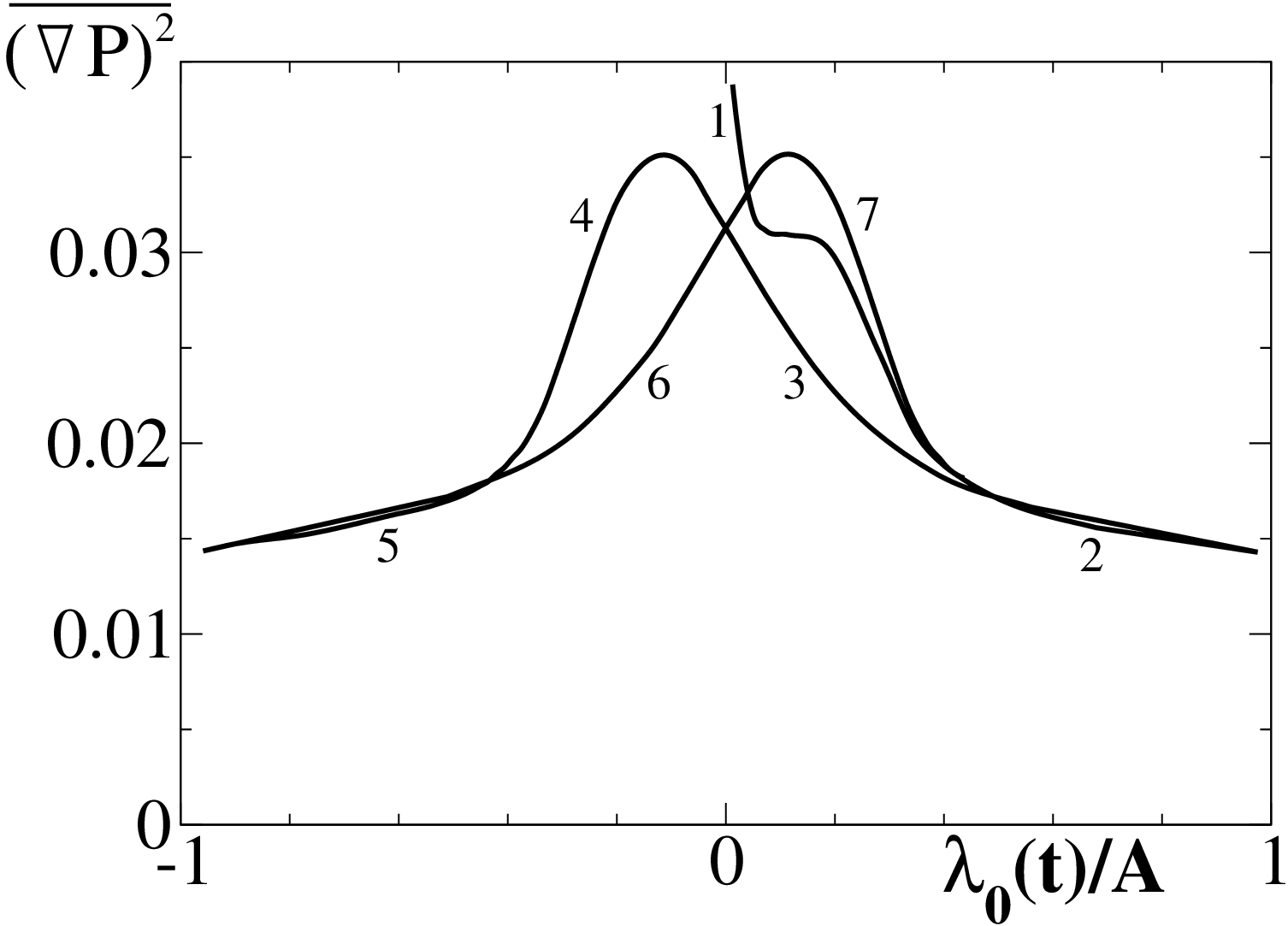}
\end{center}
\caption{The mean squared gradient of the polarization
$\overline{(\nabla P)^2}$ vs.\ $\lambda_0(t)/A$
calculated numerically for $N=3$ Fourier modes for the case of
a homogeneous external field.  The pieces of the hysteresis loop are
numbered in order of increasing time.
}
\label{grad^2}
\end{figure}

In contrast to the spatially homogeneous approximation with $N=1$,
inclusion of the non-zero modes allows us to evaluate not only the mean
polarization, but also the mean squared gradient of the polarization
(\ref{eq:grad}), which is a measure of the spatial inhomogeneity of
$P(x,t)$.  The hysteresis loop for this quantity, calculated for $N=3$
for the homogeneous external field, is shown in Fig.~\ref{grad^2}.
An interesting feature is that $\overline{{(\nabla P)}^2}$ has maxima at
times which roughly correspond to those at which the fastest switching of
the mean polarization takes place.  This provides evidence that the
polarization switching quite often, with a remarkable probability, is
accompanied by formation of a spatially inhomogeneous structure.
The insertions in Fig.~\ref{hist} illustrate such a scenario.
By including more Fourier modes, this eventually could be identified
with a domain-like structure. This is the most probable scenario
expected from the known studies of the polarization reversal 
as a nucleation and domain growth process (see, e.~g., 
\cite{KO1,KO3,CLKC07} and references therein).

\begin{figure}
\begin{center}
\includegraphics[scale=0.4]{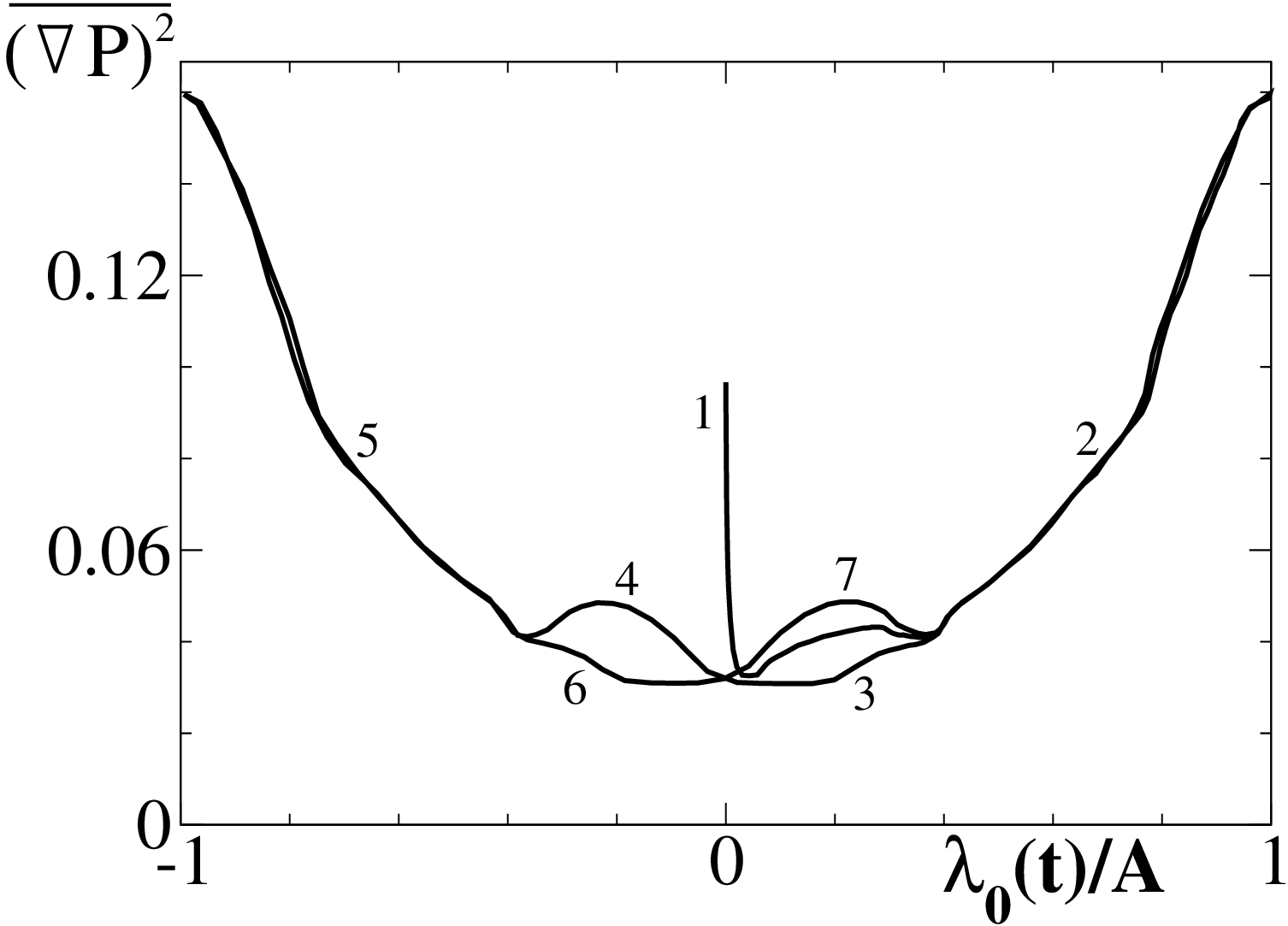}
\end{center}
\caption{The mean squared gradient of polarization
$\overline{(\nabla P)^2}$ vs $\lambda_0(t)/A$
calculated numerically for $N=3$ Fourier modes for the case of the
non-homogeneous external field.  Pieces of the hysteresis loop are
numbered in order of increasing time.
}
\label{nh_grad^2}
\end{figure}

In Fig.~\ref{nh_grad^2}, the $\overline{{(\nabla P)}^2}$ hysteresis loop
is plotted for the non-homogeneous external field.  The maxima related to
the polarization switching are observed in this case too.  Contrary to the
case for the homogeneous field, which tends to order the polarization in one
direction, the sign alternating non-homogeneous field forces the
formation of a non-uniform spatial distribution in the
polarization $P(x,t)$.
It can be seen from Figs.~\ref{grad^2} and \ref{nh_grad^2} that the
mean squared gradient of $P(x,t)$ decreases at the largest absolute values
of the field if the field is homogeneous (Fig.~\ref{grad^2}) and, remarkably,
increases for the non-homogeneous field (Fig.~\ref{nh_grad^2}).

\section{Conclusions}

In the present work a multi-dimensional Fokker-Planck equation has
been derived which describes the kinetics of polarization
switching in a ferroelectric in the presence of an external field.
The probability distribution function for this equation depends on
a set of Fourier amplitudes.  Calculations were performed for a
spatially homogeneous approximation retaining only the zero mode
${\bf k}={\bf 0}$, as well as for three Fourier modes
($k=-k_L,0,k_L$) for both homogeneous and non-homogeneous external
fields.  The hysteresis of the mean polarization $\bar P$ and of
the mean squared gradient of the polarization $\overline{(\nabla
P)^2}$ were calculated and compared. In particular, it was found
that the $\bar P$ hysteresis loop becomes slightly narrower when
more Fourier modes are included.  This loop is qualitatively the
same as that observed in real ferroelectrics.  The mean squared
gradient of polarization is a measure of its inhomogeneity. The
hysteresis loop of $\overline{(\nabla P)^2}$ provides evidence
that polarization switching is accompanied by an increased spatial
inhomogeneity of $P(x,t)$, as expected from
the known theoretical approaches
describing the polarization reversal as a nucleation and growth
of domains.

Although only a few Fourier modes were included in the actual
calculations, the problem was treated non-perturbatively and without
the mean-field approximation.  More precisely, our calculations were based
on exact equations for the probability distribution function in the case
for which the Hamiltonian (\ref{eq:HF}) contains a given number of modes.
The present results thus can be used to test various possible approximations.

\section{Acknowledgements}

 This work was carried out under the HPC-EUROPA project
(RII3-CT-2003-506079), with the support of the European Community
- Research Infrastructure Action under the FP6 "Structuring the
European Research Area" Programme.


\begin{thebibliography}{}

\bibitem{Ma} Shang-Keng Ma, \textit{Modern Theory of Critical
Phenomena}, W.A.~Benjamin, Inc., New York, 1976

\bibitem{Justin} J.~Zinn-Justin, \textit{Quantum Field Theory and
Critical Phenomena}, Clarendon Press, Oxford, 1996

\bibitem{K} J.~Kaupu\v{z}s, Ann.~Phys.~(Leipzig) {\bf 10}, 299 (2001)

\bibitem{Parisi} G.~Parisi, N.~Sourlas, Phys.\ Rev.\ Lett.\
{\bf 43}, 744 (1979)

\bibitem{Haken} H.~Haken, \textit{Synergetics}, Springer-Verlag, Berlin/
Heidelberg/New York, 1978

\bibitem{Liu} J-M. Liu, Q. C. Li, W. M. Wang, X. Y. Chen, G. H. Cao, X.
H. Liu, Z. G. Liu, J. Phys.: Condens.\ Matter {\bf 13}, L153, 2001

\bibitem{Talkner} P. Talkner, N. J. Phys.\ {\bf 1}, 4.1--4.25, 1999

\bibitem{Wan} Y. Shih, Wei-Heng Shih, Ilhan A. Aksay, Phys.\ Rev.\ B
{\bf 50}, 15575 (1994)

\bibitem{Shiino} M. Shiino, Phys.\ Rev.\ A {\bf 36}, 2393 (1986)

\bibitem{K1} J. Kaupu\v{z}s, Phys.\ Stat.\ Sol.\ {\bf 195}, 325 (1996)

\bibitem{K2} J. Kaupu\v{z}s, E. Klotins, Ferroelectrics {\bf 296}, 239 (2003)

\bibitem{DM} N. Drozdov, M. Morillo, Phys.\ Rev.\ E {\bf 54}, 3304 (1995)

\bibitem{Klotins} E. Klotins, Physica A {\bf 340}, 196 (2004)

\bibitem{KK} E. Klotins, J. Kaupu\v{z}s, Journal of European Ceramical
Society {\bf 25}, 2553 (2005)

\bibitem{KO1} S. A. Kukushkin, A. V. Osipov, Phys. Rev. B {\bf 65}, 174101 (2002)

\bibitem{KO2} S. A. Kukushkin, A. V. Osipov, Physics of the Solid State {\bf 43}, 82 (2001)

\bibitem{KO3} S. A. Kukushkin, A. V. Osipov, Physics of the Solid State {\bf 43}, 90 (2001)

\bibitem{CLKC05} S. Choudhury, Y. L. Li, C. E. Krill, L. Q. Chen, Acta Materialia {\bf 53}, 5313 (2005)

\bibitem{CLKC07} S. Choudhury, Y. L. Li, C. E. Krill, L. Q. Chen, Acta Materialia {\bf 55}, 1415 (2007)

\bibitem{ZB1} W. Zhang, K. Bhattacharya, Acta Materialia {\bf 53}, 185 (2005)

\bibitem{ZB2} W. Zhang, K. Bhattacharya, Acta Materialia {\bf 53}, 199 (2005)

\bibitem{IMAMSY} M. Iwata, T. Morishita, R. Aoyagi, M. Maeda, I. Suzuki, N. Yasuda,
Ferroelectrics {\bf 355}, 28 (2007)

\bibitem{MMSYU} E. Milov, V. Milov, B. Strukov, K. Ymazaki, Y. Uesu,
Ferroelectrics {\bf 341}, 39 (2006)


\bibitem{PolskyRimshans1986} B. S. Polsky, J. S. Rimshans,
Solid State Electron. {\bf 29}, 321 (1986)

\bibitem{KaupRim2005} J. Kaupuzs, J. Rimshans, N. Smyth,
In: "Electromagnetic Fields in Mechatronics, Electrical and
Electronic Engineering", Proc. of ISEF'05, IOS Press, 58 (2006).

\end{thebibliography}
\end{document}